\documentclass[12pt]{article}
\pdfoutput=1
\usepackage{fullpage}
\usepackage{graphicx}
\usepackage{amssymb}
\usepackage{amsmath}
\usepackage{multicol}
\usepackage{ulem}
\usepackage{hyperref}
\usepackage{caption}
\usepackage{subcaption}
\usepackage{float}

\newcommand{\nh}{non-holomorphic }
\newcommand{\pr}{\partial{}}

\newcommand{\bx}{\mbox{\boldmath $x$}}

\newcommand{\bn}{\mbox{\boldmath $n$}}

\newcommand{\me}{minimum-energy }

\date{}
\usepackage{wrapfig}
\usepackage[margin=2.50cm]{geometry}

\title{Negative Baryon density and the Folding structure of the B=3 Skyrmion}
\author{David Foster\footnote{df211@kent.ac.uk} ~and Steffen Krusch\footnote{S.Krusch@kent.ac.uk} \\  
{\it School of Mathematics, Statistics and Actuarial Science,}\\
{\it University of Kent, Canterbury CT2 7NF, United Kingdom} \\
}

\begin{document}

\maketitle

\begin{abstract}
The Skyrme model is a non-linear field theory whose solitonic solutions, once quantised, describe atomic nuclei. The classical static soliton solutions, so-called Skyrmions, have interesting discrete symmetries and can only be calculated numerically. Mathematically, these Skyrmions can be viewed as maps between to two three-manifolds and, as such, their stable singularities can only be folds, cusps and swallowtails. Physically, the occurrence of singularities is related to negative baryon density. In this paper, we calculate the charge three Skyrmion to a high resolution in order to examine its singularity structure in detail. Thereby, we explore regions of negative baryon density. We also discuss how the negative baryon density depends on the pion mass.
\end{abstract}

\section{Introduction}
The Skyrme model is a $(3+1)$-dimensional nonlinear theory of pion’s and was conjectured by Skyrme as a model of baryons \cite{Skyrme:1961vq}. Subsequently, Witten \cite{Witten:1983tw} derived it as an effective action of QCD -- in the large colour limit. The \me solutions of the model are called Skyrmions and are identified as baryons. In this article we are interested in static solutions of the Skyrme model, best defined by the static energy functional
\begin{equation}
E = \frac{1}{12 \pi^2}\int \left\{ -\frac{1}{2}\mbox{Tr}(R_i R_i) -\frac{1}{16} \mbox{Tr}([R_i,R_j][R_i,R_j]) + m^2\mbox{Tr}(I_{2}-U) \right\} d^3 x, \label{Skyrme-eng}
\end{equation}
where $R_\mu$ is an $su(2)$-valued current, $R_\mu = (\pr_\mu U)U^{\dag}$ and $U({\bf x},t)$ is an $SU(2)$-valued scalar field. The parameter $m$ is related to the physical pion mass $m_\pi$ via $m=2 m_\pi/(e F_\pi)$ where $F_\pi$ is the pion decay constant and $e$ is the Skyrme constant. Here we have presented the energy functional in so-called Skyrme-units, see \cite{Adkins:1983ya, Adkins:1983hy} for the ``standard values'' and \cite{Battye:2005nx,Manton:2006tq} for a more detailed discussion of our current understanding. For non-zero pion mass the finite energy requirement forces the field $U:\mathbb{R}^3 \to \mbox{SU}(2)$ to satisfy
$$
\lim_{r \to \infty} U = I_{2},
$$
and we choose the same boundary condition for $m_\pi = 0.$  
This one-point compactifies $\mathbb{R}^3$ to the 3-sphere $S^3$. Using the fact that the group manifold of $\mbox{SU}(2)$ is also $S^3$ the field can be extended to a map $U:S^3 \to S^3$. All finite energy field configurations $U({\bf x})$ belong to an element of $\pi_3(S^3)=\mathbb{Z}$, and hence have an associated integer $B\in \mathbb{Z}$. It is this $B$ which is identified as the baryon number and can be explicitly calculated as
\begin{eqnarray}\label{Skyrme-B}
 B&=& -\frac{1}{24\pi^2} \int \varepsilon_{ijk} \mbox{Tr}(R_iR_jR_k) d^3x, \\ 
&=& \frac{1}{2 \pi^2} \int \det\left(J({\bf x})\right) d^3x, \nonumber
\end{eqnarray}
where $J({\bf x})$ is the Jacobian of the map \cite{Manton:1987xt}. Naively, the density of $B$ could be assumed to be positive for all ${\bf x}\in \mathbb{R}^3$, but this does not have to be the case. It was shown in \cite{Houghton:2001fe} that  as $\mathbb{R}^3$ covers $S^3$ -- the group manifold of $SU(2)$ -- there is a folding structure. On these folds the Jacobian-determinant becomes zero which results in tubes of zero baryon density. Inside these tubes singularity theory predicts regions of negative baryon density, but this was never observed in numerical calculations. This is what we are interested in this paper. It is useful to define two quantities,
\begin{eqnarray}
{\cal B}_+(x) &=& \frac{1}{2\pi^2}\max\{\det (J({\bf x})),0\}, \\ \nonumber
{\cal B}_-(x) &=& \frac{1}{2\pi^2}\max\{-\det (J({\bf x})),0\}, \\ \nonumber
\end{eqnarray}
where $B=\int {\cal B}(\bx) d^3x$ and trivially ${\cal B}({\bf x})= {\cal B}_+({\bf x})-{\cal B}_-({\bf x})$. We refer to these quantities as the positive baryon density and negative baryon density, respectively.

The paper is laid out as follows. Section \ref{geometry} reviews the geometric formulation of the Skyrme model proposed in \cite{Manton:1987xt} and discusses the Jacobian. In particular, we show an important identity relating the Jacobian of the Skyrme field to a simpler quantity. Section \ref{ratmapansatz} reviews the rational map ansatz \cite{Houghton:1997kg} as well as the \nh rational map ansatz \cite{Houghton:2001fe}. In section \ref{results} we examine the folding structure of the $B=3$ Skyrmion using both full field simulation and the \nh rational map ansatz. Section \ref{pionmass} discusses the effects of the pion mass term on the folding structure. In section \ref{expansion} we locally expand the pion fields around the origin and reproduce the singular surface. We end with a conclusion.

\section{Geometric setting for Skyrmions}
\label{geometry}
Instead of considering the Skyrme model in a physical field theoretic setting, it can be very insightful to consider it geometrically \cite{Manton:1987xt, Krusch:2000gb}. In the following, we give a brief account of this approach and set up our notation. 

A field configuration is a map ${\pmb \pi}$ from a physical space $\mathbb{R}^3$
to a target space $SU(2)$. Physical space $\mathbb{R}^3$ and target space $SU(2) \cong S^3$ both are $3$-dimensional, connected and orientable Riemannian manifolds. Here we choose the Kronecker delta $\delta_{ij}$ as the flat-metric on $\mathbb{R}^3$ and $\gamma_{\alpha \beta}$ as the metric  on $S^3$. We denote a point in $\mathbb{R}^3$ as ${\bf x}$ and a point in $S^3$, the image of ${\bf x}$, as ${\pmb \pi}({\bf x}).$
As $\mathbb{R}^3$ is flat we trivially choose its dreibein to be $\delta_{ij}$, and we choose ${\zeta_\mu}^\alpha({\pmb \pi}({\bf x}))$ as the dreibein on $S^3$ with 
\begin{equation}
{\zeta_\mu}^\alpha {\zeta_\nu}^\beta
\gamma_{\alpha \beta} = \delta_{\mu  \nu}.
\end{equation}
Note the inverse of ${\zeta_\mu}^\alpha$ is ${\zeta^\mu}_\alpha,$
i.e. ${\zeta^\mu}_\alpha {\zeta_\nu}^\alpha = {\delta^\mu}_\nu$ and  
${\zeta^\mu}_\alpha {\zeta_\mu}^\beta = {\delta_\alpha}^\beta.$

Now we can define the Jacobian matrix associated with the map 
${\pmb \pi}({\bf x})$ as
\begin{equation}
{J_i}^\mu ({\bf x}) =  (\partial_i \pi^\alpha({\bf x})) {\zeta^\mu}_
\alpha ({\pmb \pi}({\bf x})).
\end{equation}
The matrix ${J_m}^\mu ({\bf x})$ is a measure of the deformation induced by
the map ${\pmb \pi}$ at the point ${\bf x}$ in $\mathbb{R}^3$. We can define a useful quantity called the strain tensor $D_{ij}$, as
\begin{equation}
\label{Dmn}
D_{ij}({\bf x}) = {J_i}^\mu ({\bf x}){J_j}^\nu ({\bf x}) \delta_{\mu \nu}
= (\partial_i \pi^\alpha({\bf x}))(\partial_j \pi^\beta({\bf x})) \gamma_{\alpha \beta}.
\end{equation}
The strain tensor is invariant under rotations in target space (i.e. rotations of the frame fields ${\zeta_\mu}^\alpha({\pmb \pi}({\bf x}))$), but not under  orthogonal rotations of the physical space $\mathbb{R}^3$. But it is well known that the characteristic polynomial $P=\mbox{det}(D-\lambda I_{3})$ is invariant under rotations. So, we can now define the three invariants,
\begin{eqnarray}
\mbox{Tr}(D) &=& \lambda_1^2+\lambda_2^2+\lambda_3^2, \\ \nonumber
\frac{1}{2}(\mbox{Tr}(D))^2 - \frac{1}{2} \mbox{Tr}(D^2) &=&\lambda_1^2\lambda_2^2 + \lambda_1^2 \lambda_3^2 + \lambda_2^2 \lambda_3^2, \\ \nonumber
\mbox{det(D)}&=&\lambda_1^2 \lambda_2^2 \lambda_3^2, \nonumber
\end{eqnarray}
where $\lambda_1^2,\lambda_2^2,\lambda_3^2$ are the non-negative eigenvalues of the symmetric matrix $D_{ij}$.

Reparametrising the $SU(2)$-valued field in the traditional way with the three Pauli matrices, $\tau_a$, and the four scalar fields $\sigma({\bf x}), \pi_a ({\bf x}) ~ (a=1,2,3)$ as
\begin{equation}
\label{U(x)}
U({\bf x}) = \sigma({\bf x}) +i {\pmb \pi}({\bf x}) \cdot {\pmb \tau},
\end{equation}
we see that $\sigma({\bf x})$ and ${\pmb \pi}({\bf x})$ must satisfy the constraint $\sigma^2+{\pmb \pi} \cdot {\pmb \pi}= 1$ for all  ${\bf x} \in {\mathbb R}^3$. As pointed out in \cite{Krusch:2000gb} $\pi^\alpha({\bf x})$ in the above geometric discussion can be identified with the vector $\pi_a$ in \eqref{U(x)}  and $\sigma({\bf x})$ is a function of $\pi_a({\bf x})$ which ensures that $U({\bf x})\in SU(2)$. Now, we can define the induced-metric on $S^3$ as
$$
\gamma_{\alpha \beta}({\bf x})=\delta_{\alpha \beta} + \frac{\pi_\alpha \pi_\beta}{\sigma^2}.
$$
A short calculation in \cite{Krusch:2000gb} shows that
$$
D_{ij}({\bf x}) = -\frac{1}{2} \mbox{Tr}(R_i R_j),
$$
and the energy functional \eqref{Skyrme-eng} can be rewritten as
\begin{eqnarray}
\label{ED}
E&=& \mbox{Tr}(D) +\frac{1}{2}(\mbox{Tr}(D))^2-\frac{1}{2}\mbox{Tr}(D^2), \\ \nonumber
&=&\lambda_1^2+\lambda_2^2+\lambda_3^2+\lambda_1^2\lambda_2^2 + \lambda_1^2 \lambda_3^2 + \lambda_2^2 \lambda_3^2.
\end{eqnarray}
Also, it is easy to see from the above relations that the baryon number integral \eqref{Skyrme-B} can be written in terms of these eigenvalues as
\begin{equation}
B=\frac{1}{2\pi^2} \int \lambda_1 \lambda_2 \lambda_3 ~ d^3x.
\end{equation}

\subsection{The Jacobian}

In the following we calculate the Jacobian in stereographic coordinates. This is needed for the analysis of the numerically found solutions.
Consider the Skyrme field in sigma model coordinates $(\sigma({\bf
  x}), \pi_i({\bf x}))$ with $\sigma^2+ {\pmb \pi}^2 = 1.$
Then we can define stereographic coordinates by projecting from the
North pole $N$ $(\sigma=1)$ as
\begin{equation}
\Phi_{N}^\alpha = \frac{\pi^\alpha}{1-\sigma}.
\end{equation}
Note that the metric in stereographic coordinates for this chart is given by
\begin{equation}
\label{gStereo}
ds_N^2 = \sum\limits_{\alpha=1}^3 
\frac{4}{(1+{R_N}^2)^2} \left(d\Phi_N^\alpha\right)^2,
\end{equation}
where ${R_N}^2 = \sum\limits_{\alpha=1}^3 \left(\Phi_N^\alpha\right)^2
= \frac{1+\sigma}{1-\sigma}.$
Since the metric (\ref{gStereo}) is diagonal we can define the frame fields
\begin{equation}
{{\zeta_N}_\mu}^\alpha = \frac{1+{R_N}^2}{2} {\delta_\mu}^\alpha\quad {\rm
  and~its~inverse}\quad 
{{\zeta_N}^\mu}_\alpha = \frac{2}{1+{R_N}^2} {\delta^\mu}_\alpha.
\end{equation}
Hence, the Jacobian is given by
\begin{equation}
{{J_N}_m}^\mu = 
\frac{\partial \Phi_N^\alpha}{\partial x_m} {{\zeta_N}^\mu}_\alpha
= \frac{\partial \Phi_N^\mu}{\partial x_m}  \frac{2}{1+{R_N}^2}.
\end{equation}
Using (\ref{Dmn}) we can calculate the strain tensor $D_{mn}$ and obtain
\begin{equation}
\label{DN}
{D_N}_{mn} = \frac{4}{(1+{R_N}^2)^2} 
\frac{\partial \Phi_N^\alpha}{\partial x_m}
\frac{\partial \Phi_N^\beta}{\partial x_n} \delta_{\alpha \beta}.
\end{equation}
Note \lq{}$N$\rq{} is not an index to be summed over.

To obtain a well-defined $SO(3)$ frame bundle the frame fields $\zeta_S$ have to be chosen as
\begin{equation}
{{\zeta_S}_\mu}^\alpha = -\frac{1+{R_S}^2}{2} {\delta_\mu}^\alpha\quad {\rm
  and~its~inverse}.
\end{equation}
See e.g. \cite{Krusch:2003xh} for further details.
This leads to
\begin{equation}
\label{JS}
{{J_S}_m}^\mu = 
 -\frac{\partial \Phi_S^\mu}{\partial x_m}  \frac{2}{1+{R_S}^2},
\end{equation}
where $\Phi_{S}^\alpha = \frac{\pi^\alpha}{1+\sigma}$ is the South pole projection. \\
The minus sign in (\ref{JS}) arises because stereographic coordinates
are related by inversion, which has negative determinant.  The expression for the 
strain tensor ${D_S}_{mn}$ is
\begin{equation}
\label{DN}
{D_S}_{mn} = \frac{4}{(1+{R_S}^2)^2} 
\frac{\partial \Phi_S^\alpha}{\partial x_m}
\frac{\partial \Phi_S^\beta}{\partial x_n} \delta_{\alpha \beta} \nonumber
\end{equation}

There is a sign ambiguity when defining the Jacobians $J_N$ and
$J_S$. Here we have chosen the Jacobian such that the standard
hedgehog has positive Jacobian. In fact, for $B=1$ the
hedgehog can be written as
\begin{equation}
\sigma = \cos (f(r)) \quad {\rm and} \quad 
\pi_i = \frac{x_i}{r} \sin (f(r)),
\end{equation}
where $r = \sqrt{x_1^2+x_2^2+x_3^2}$ and $f$ is a real radial shape function
that can be determined numerically. Asymptotically, 
$f(r) \approx \pi - A r$ as $r \to 0$ with $A >0$ and $f(r) \approx
\frac{C}{r^2}$ as $r \to \infty$ with $C\approx 2.16,$ see e.g. \cite{Manton:2004tk}. 
Near the origin, $\sigma  \approx -1,$ so we project from the north
pole and obtain $\det J_N \approx A^3 >0,$ whereas as $r\to \infty,$
$\sigma \approx 1$ we project from the south pole and $J_S \approx
\frac{2C^3}{r^9} >0.$  
These expression of the Jacobians ($J_N$ and $J_S$) 
are useful when examining the
behaviour near the north and south pole, which corresponds to the
vacuum and the anti-vacuum respectively.

For analysis of numerical data later in the paper, we can express the Jacobian directly in terms of the four pion fields as
\begin{eqnarray}
\left({{J_N}_m}^\mu\right) &=& (1-\sigma) \frac{\partial
  {\Phi_N}^\mu}{\partial x_m}\\ 
&=&  \frac{\partial \pi^\mu}{\partial x_m} + 
\frac{\pi^\mu}{(1-\sigma)} \frac{\partial \sigma}{\partial x_m}\\
&=& \frac{\partial \pi^\mu}{\partial x_m} - 
\sum\limits_\nu
\frac{\pi^\mu \pi^\nu}{\sigma (1-\sigma)} \frac{\partial \pi^\nu}{\partial x_m},
\end{eqnarray}
where we used the identity
\begin{equation}
\label{ID}
\sigma \frac{\partial \sigma}{\partial x_j} = -\sum\limits_{\nu=1}^3
\pi^\nu\frac{\partial \pi^\nu}{\partial x_j},
\end{equation}
which is derived by differentiating the normalisation condition.

As a check, we can evaluate 
\begin{equation}
D_{mn} = {J_m}^\mu (x){J_n}^\nu (x) \delta_{\mu \nu}
\end{equation}
and obtain 
\begin{equation}
D_{mn}=\frac{\partial \pi^\mu}{\partial x_m} \frac{\partial \pi^\nu}{\partial
  x_n}
\left(\delta_{\mu \nu} - \frac{\pi_\mu \pi_\nu}{\sigma^2}\right) .
\end{equation}
The term in the brackets corresponds to the induced metric in
$(\sigma, \pi_i)$ coordinates. Using the identity (\ref{ID}) we obtain
the more familiar expression
\begin{equation}
D_{mn} = 
\frac{\partial \sigma}{\partial x_m} \frac{\partial \sigma}{\partial
  x_n}
+ \frac{\partial \pi^\mu}{\partial x_m} \frac{\partial \pi^\nu}{\partial x_n}\delta_{\mu \nu}.
\end{equation}
With this expression, equation (\ref{ED}) gives the pion field version of the Skyrme Lagrangian which is often used for numerical simulations, see e.g. \cite{Battye:2001qn}. 
It can be shown by direct calculation that 
$\det(J_N) = -\det (J_0)/\sigma$ where
\begin{equation}
J_0 = \frac{\partial \pi_i}{\partial x_j}.
\end{equation}
Hence $\det J_N$ vanishes when $\det J_0$ does for $\sigma \neq 0,$ therefore $J_N$ and $J_0$ have the same singular surfaces.

\section{Rational map ansatz}
\label{ratmapansatz}
The rational map ansatz \cite{Houghton:1997kg} makes use of the feature that a point in $\mathbb{R}^3$ can be written in polar coordinates $(r,z)$ where the angular coordinate $z$ is represented by a point on the Riemann sphere via $z = {\rm e}^{i\phi} \tan \frac{\theta}{2}.$ The ansatz then takes the form
\begin{equation}
\label{ratmap}
U(r,z)=\exp(i f(r) \bn(z) \cdot \boldsymbol{\tau}),
\end{equation}
where $f(r)$ is a real valued profile function with the boundaries $f(0)=\pi$ and $f(\infty)=0$; $\boldsymbol{\tau}=(\tau_1,\tau_2,\tau_3)$ is the triplet of Pauli matrices and $\bn$ is the unit vector 
$$
\bn(z) = \frac{1}{1+|R(z)|^2}(R(z)+\overline{R(z)},i(\overline{R(z)}-R(z)),1-|R(z)|^2).
$$
$R(z)$ is a holomorphic rational map between Riemann spheres, and is given by the two polynomials $p(z),q(z)$,
$$
R(z) =\frac{p(z)}{q(z)}.
$$
Then the baryon number is equal to the algebraic degree of $R(z)$,
$$ 
\deg\left(R(z)\right)=\max\left\{\deg(p(z)),\deg(q(z))\right\}.
$$ 
Such an ansatz is a suspension and gives rise to an isomorphism between $\pi_3 (S^3)$ and $\pi_2(S^2)$. It is the choice of $R(z)$ which replicates the polyhedral shape of Skyrmions, and the symmetries of the accepted numerical solutions. The benefit of the rational map ansatz is that it  gives rise to the following three simple eigenvalues of the strain tensor,
\begin{eqnarray}
\lambda_1&=&-f\rq{}(r) \\ \nonumber
\lambda_2=\lambda_3&=&\frac{\sin f}{r} \frac{1+|z|^2}{1+|R|^2} \left|\frac{dR}{dz} \right|.
\end{eqnarray}
This now gives the simple radial energy functional
\begin{equation}
E=4\pi \int \left(f\rq{}^2 r^2+2B(f\rq{}^2 +1)\sin^2 f+\mathcal{I}\frac{\sin^4f}{r^2}+2m^2 r (1-\cos f) \right)dr, \label{E-rat-map}
\end{equation}
where
\begin{equation}
\mathcal{I}=\frac{1}{4 \pi} \int \left( \frac{1+|z|^2}{1+|R|^2} \left|\frac{dR}{dz}\right|\right)^4 \frac{2 i dz d\bar{z}}{(1+|z|^2)^2}.
\end{equation}
This energy functional can be easily minimised by choosing the correct degree polynomials $p(z),q(z)$ which minimise $\mathcal{I}$, then numerically minimise the profile function $f(r)$. The minimum energy solutions found using this method only exceed the non-symmetry numerical solutions by about $3\%,$ see e.g. \cite{Houghton:1997kg, Battye:2001qn, Battye:2002wc}. 

This holomorphic ansatz is very successful at capturing the major features of the Skyrme model and is a very useful technique to give initial configurations which are close to the minimum energy solutions for numerical minimisation. This avoids a numerically expensive collisions which were previously used to create appropriate initial conditions \cite{Battye:1997qq, Battye:2001qn}. A feature of the holomorphic ansatz is that it locally preserves orientation. Hence there cannot be regions of negative baryon density. This constrains the possible configurations. We can extend the ansatz to \nh $R(z,\bar{z})$ as in \cite{Houghton:2001fe}. This now allows points in $\mathbb{R}^3$ with negative baryon density. In this case the three eigenvalues of the strain tensor $D_{ij}$ are slightly more complicated and are
\begin{eqnarray}
\lambda_1 &=& -f\rq{}(r), \\ \nonumber
\lambda_2 &=& \frac{\sin f}{r} (|R_z| + |R_{\bar{z}}|)\frac{1+|z|^2}{1+|R|^2}, \\ \nonumber
\lambda_3 &=& \frac{\sin f}{r} (|R_z| - |R_{\bar{z}}|)\frac{1+|z|^2}{1+|R|^2}, \\ \nonumber
\end{eqnarray}
where $R_z$ and $R_{\bar{z}}$ are the derivatives of $R$ with respect to $z$ and ${\bar{z}},$ repectively.
It is now apparent that for \nh  (or non-antiholomorphic) $R$ the angular strains $\lambda_2,\lambda_3$ are no longer isotropic, and only when $R_z$ (or $R_{\bar{z}}$) equal zero do we regain the previous holomorphic ansatz. The energy functional for this more general ansatz is
\begin{equation}
E=\frac{1}{3\pi} \int \left(f\rq{}^2 r^2+2\mathcal{J}(f\rq{}^2 +1)\sin^2 f+\tilde{\mathcal{I}}\frac{\sin^4f}{r^2} +2m^2 r (1-\cos f)\right)dr, \label{nonholo-E-rat-map}
\end{equation}
where
\begin{eqnarray}
\mathcal{J} &=&\frac{1}{4\pi} \int \left((|R_z|^2 + |R_{\bar{z}}|^2) \left(\frac{1+|z|^2}{1+|R^2} \right)^2\right)\frac{2 i dz d\bar{z}}{(1+|z|^2)^2}, \\ \label{nonholo-J}
\tilde{\mathcal{I}}&=& \frac{1}{4\pi} \int \left((|R_z|^2 - |R_{\bar{z}}|^2)\left(\frac{1+|z|^2}{1+|R|^2}\right)^2\right)^2\frac{2 i dz d\bar{z}}{(1+|z|^2)^2}, \label{nonholo-I}
\end{eqnarray}
and 
\begin{equation}
B= \frac{1}{2 \pi} \int f\rq{}(r) \sin^2f(r) \left((|R_z|^2 - |R_{\bar{z}}|^2)\left(\frac{1+|z|^2}{1+|R|^2}\right)^2\right)\frac{2 i dz d\bar{z}}{(1+|z|^2)^2}. \label{nonholo-B}
\end{equation}
Again $f(r)$ is a profile function which is a solution of the ODE
\begin{equation}
(r^2+2\mathcal{J} \sin^2 f)f\rq{}\rq{} +2rf\rq{}+(\mathcal{J}f\rq{}^2-\mathcal{J}-\tilde{\mathcal{I}}\frac{\sin^2f}{r^2})\sin 2 f-m^2 r \sin f =0, \label{non-holo ode}
\end{equation}
with the boundary conditions $f(0)=\pi$ and $f(\infty)=0$. 

As above, if we restrict to maps of the form 
$$
R(z,\bar{z})=\frac{p(z,\bar{z})}{q(z,\bar{z})},
$$
where the ${\bar z}$ dependence is chosen in such a way as to preserve the symmetry of the original minimal energy rational map.
Then the baryon number is generically equal to $N_1-N_2$, where $N_1$ is the maximal holomorphic degree of $(p,q)$ and $N_2$ is the maximal antiholomorphic degree of $(p,q)$. 

In \eqref{non-holo ode} ${\cal J}$ replaces the role of $B$ in the holomorphic rational map ansatz. This has a significant effect on the profile function $f$. To understand this we linearise \eqref{non-holo ode} about $r=0$ and set $f(r) = \pi - \nu(r)$ where $\nu(r) \ll 1.$
This gives the new linear ODE
\begin{equation}
\label{nu}
r^2 \nu\rq{}\rq{}+2r\nu\rq{}-2{\cal J}\nu+m^2r\nu=0,
\end{equation}
with the solution
\begin{equation}
\label{f(r)=0}
f(r) = 
 \pi - Cr^{\left( \frac{-1+\sqrt{1+8{\cal J}}}{2}\right)}.
\end{equation}
For $m=0,$ equation (\ref{nu}) is a Cauchy-Euler equation, whereas for $m\neq 0$ the solution is given in terms of a Bessel function. Numerical calculations show that as the pion mass is increased, ${\cal J}$ becomes larger and hence $f$ decays slower about the origin. We can now play the same trick for the limit of $r$ going to infinity. Here we set $f(r)=\epsilon(r)$ where $\epsilon(r) \ll 1,$
giving the linearised equation
$$
r^2 \epsilon\rq{}\rq{} +2r\epsilon\rq{}-2{\cal J}\epsilon -m^2 r\epsilon=0.
$$
This gives the solutions
$$
f(r) = \begin{cases} C r^{-\left(\frac{1+\sqrt{1+8{\cal J}}}{2}\right)}, & \mbox{if } m=0, \\
 \frac{C {\rm e}^{-2m\sqrt{r}}}{r^{3/4} } \left(1+\frac{4(1+8{\cal J})-1}{16m\sqrt{r}} + ...\right), & \mbox{if } m \neq 0. \end{cases}
$$
The effect of the mass term is to make the solution more localised around the origin.

\section{The $B=3$ Skyrmion}
\label{results}
It has been known for a long time that the baryon density for the $B=3$ Skyrmion is tetrahedrally symmetric \cite{Braaten:1989rg, Battye:1997qq}. A lot of its features can be explained by a tetrahedrally symmetric rational map \cite{Houghton:1997kg}. But, as shown in \cite{Houghton:2001fe}, the \nh rational map which allows negative baryon density has lower energy that the holomorphic map. The \nh rational map gives rise to four singular tubes ($\det(J)=0$) which start at the origin, pass through the faces of the tetrahedron then go off to infinity. It has been shown \cite{Houghton:2001fe} that there are three folding lines equally spaced along these tubes. Also, in the centre of these tubes there are regions of negative baryon density.  This inspired us to understand the form and amount of negative baryon density in actual \me Skyrme solutions. 

The family of rational maps for the $B=3$ Skyrmion with the correct symmetry is
\begin{equation}
R=\frac{p_1 \cos \theta +p_2 \sin \theta}{q_1 \cos \theta + q_2 \sin \theta}, \label{B=3-nonholo}
\end{equation}
where the polynomials $p_1,p_2,q_1$ and $q_2$ are
\begin{equation}
\label{nhratmap}
\begin{array}{lcl}
p_1(z,\bar{z})=i\sqrt{3}z^3\bar{z}+i\sqrt{3}z^3-z\bar{z}-1,&\quad &q_1(z,\bar{z})=z^4\bar{z}+z^3-i\sqrt{3}z^2\bar{z}-i\sqrt{3}z, \\ 
p_2(z,\bar{z})=z^4-2i\sqrt{3}z^2+1,&\quad &q_2(z,\bar{z})=-z^4\bar{z}+2i\sqrt{3}z^2\bar{z}-\bar{z}.
\end{array}
\end{equation}
With a simple numerical scanning algorithm we find that the family of rational maps in (\ref{B=3-nonholo}) attains its minimum energy for $\theta=0.154$. This gives $E/B= 1.161.$ 
Note this is lower than the minimum energy for the holomorphic ansatz $(E/B=1.184)$. This value of $\theta$ is slightly lower than that found in \cite{Houghton:2001fe}. This difference is believed to be due to numerical accuracy. 

\begin{figure}[!htb]
\centering
\includegraphics[height=3.8in]{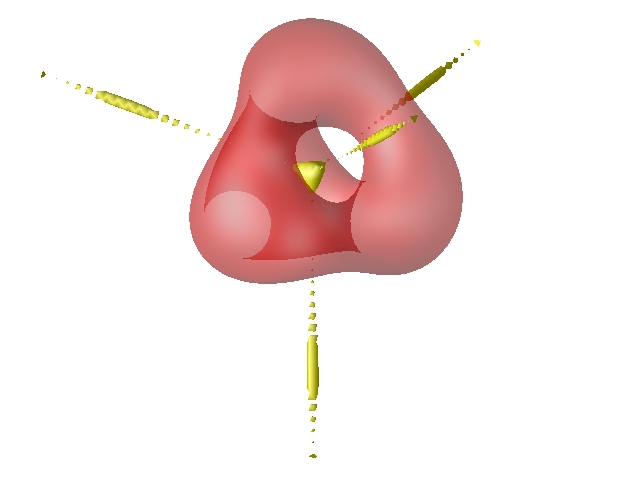} 
\caption{Baryon number three \me solution. The red surface is a level set of constant positive baryon density. The yellow surface is a level set of constant negative baryon density.}
\label{B3 negative baryon density}
\end{figure}

To find the $B=3$ minimum energy solution we numerically minimised this rational map ansatz on a lattice of $250^3$ points, with spacing $\Delta x=0.08$ using fourth order accurate derivatives. For $m=0$ we find $E/B=1.146$ and the negative baryon density $B_{-} =4.5 \times 10^{-5}$. This is in reasonable agreement with $B_{-} =9.25 \times 10^{-5}$ found for the \nh rational map for $\theta=0.154$. A surface of constant baryon density (ie. constant Jacobian determinant) is displayed in figure \ref{B3 negative baryon density} for this solution.

To truly capture these regions of negative baryon density we were forced to
have a very large box. It is known that asymptotically the $B = 3$ Skyrmion
decays as a $B = -1$ Skyrmion \cite{Manton:1994ci}. The $B=1$ Skyrmion can be calculated from the highly symmetric rational map, $R=z$, known as the hedgehog ansatz. Then the equations of motion \eqref{non-holo ode} can be linearised to obtain asymptotic behaviour of the radial profile function $f \sim C/r^2$ as $r \to \infty$ for some constant $C$. Substituting this into the radial energy-density \eqref{nonholo-E-rat-map} and baryon density \eqref{nonholo-B}, we find that
the energy-density scales as $1/r^4$ and the baryon-density as $1/r^7$.
We have checked numerically that choosing $U(\bx)=I_2$ on the boundary of the numerical box contains the baryon-density within the accuracy of our numerics. However, for an evaluation of the total energy to a high order of accuracy contributions from outside the box need to be included, see e.g. \cite{Feist:2011aa} for further details.

\begin{figure}[!htb]
       \centering
       \begin{subfigure}[b]{0.5\textwidth}
               \centering
    \includegraphics[width=\textwidth]{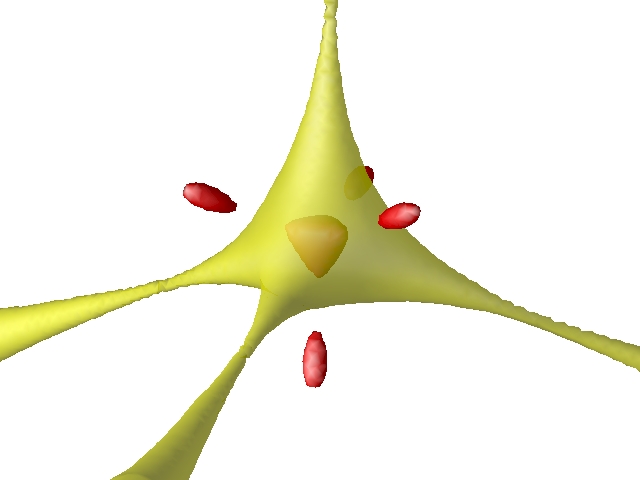}
               \caption{Yellow is a level surface of $\det(J(\bx)) \simeq 0$ and the red is surface of the anti-vacuum $\sigma \simeq -1$}
               \label{B=3-neg-tetrahedron} 
       \end{subfigure}%
        ~ 
        \begin{subfigure}[b]{0.5\textwidth}
                \centering \includegraphics[width=\textwidth]{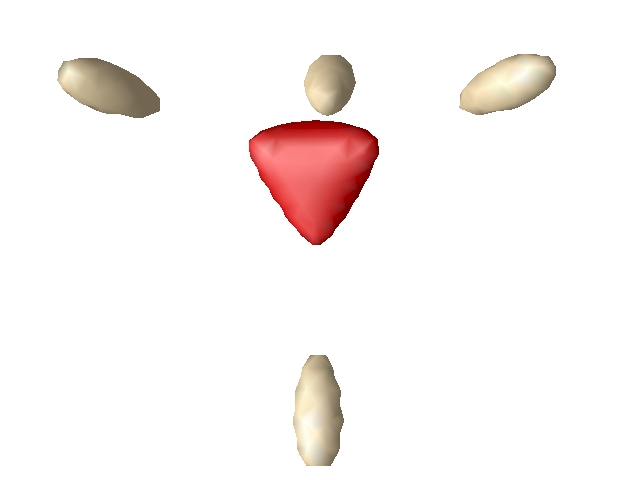}
                \caption{Preimages of the anti-vacuum. Red is where $\mbox{sign}(J(\bx))=-1$  and Ivory is where $\mbox{sign}(J(\bx))=1$.}
                \label{B=3 sigma =1}
        \end{subfigure}
        ~ 
        \caption{Minimisation of the central region of the $B=3$ \nh rational map.}\label{B=3 close to origin}
\end{figure}

Figure \ref{B3 negative baryon density} shows the regions of negative baryon density of the \me baryon number three Skyrmion. A point of particular interest is the tetrahedron of negative baryon density in the centre of the Skyrmion. Each of the corners of this dual-tetrahedron gives rise to tubes of $\det(J)=0$, which pass though the faces of the tetrahedron with positive baryon density. However,
the singular surface tubes may or may not be connected to the dual tetrahedron singular surface in the centre. With the aim of trying to understand how these singular surfaces are connected we performed a numerical simulation over a box centred at the origin with $(\Delta x = 0.01, n=180)$ sides about $9\%$ of the large box. We chose the boundary conditions, on the surface of the box, to be the rational map \eqref{B=3-nonholo}
with the \me profile function found numerically which is a good approximation to the exact solution. Our intention is to present evidence to understand the singular structure about the origin of the $B=3$ Skyrmion.  The result of our numerical calculation is presented in figure \ref{B=3 close to origin}.

Figure \ref{B=3-neg-tetrahedron} shows the folding tubes in yellow. Contrary to the conjecture in \cite{Houghton:2001fe} the tubes do not seem to pinch off to singular points. This is evidence that there are no swallowtails in the $B=3$ Skyrmion and will be discussed further in the next subsection.
Figure \ref{B=3 sigma =1} is also interesting. It shows that it is energetically favourable for the $B=3$ Skyrmion to \lq{}create\rq{} two more pre-images of the anti-vacuum, one with positive orientation and one with negative orientation. The rational map ansatz (\ref{ratmap}) has the anti-vacuum as a suspension point. Hence for the rational map ansatz there are only three pre-images of the vacuum, all at the origin and with positive orientation. These extra pre-images are very similar to the monopole zeros of the Higgs field in \cite{Sutcliffe:1996he} where they find five monopole zeros, four positive and one negative. As outlined in \cite{Sutcliffe:1996he}, one would naively suspect that a positive orientation point would annihilate with a negative orientation point. However, the configuration may be stabilised by tetrahedral symmetry. 

\begin{figure}[!htb]
      \centering
       \begin{subfigure}[b]{0.5\textwidth}
              \centering \includegraphics[width=\textwidth]{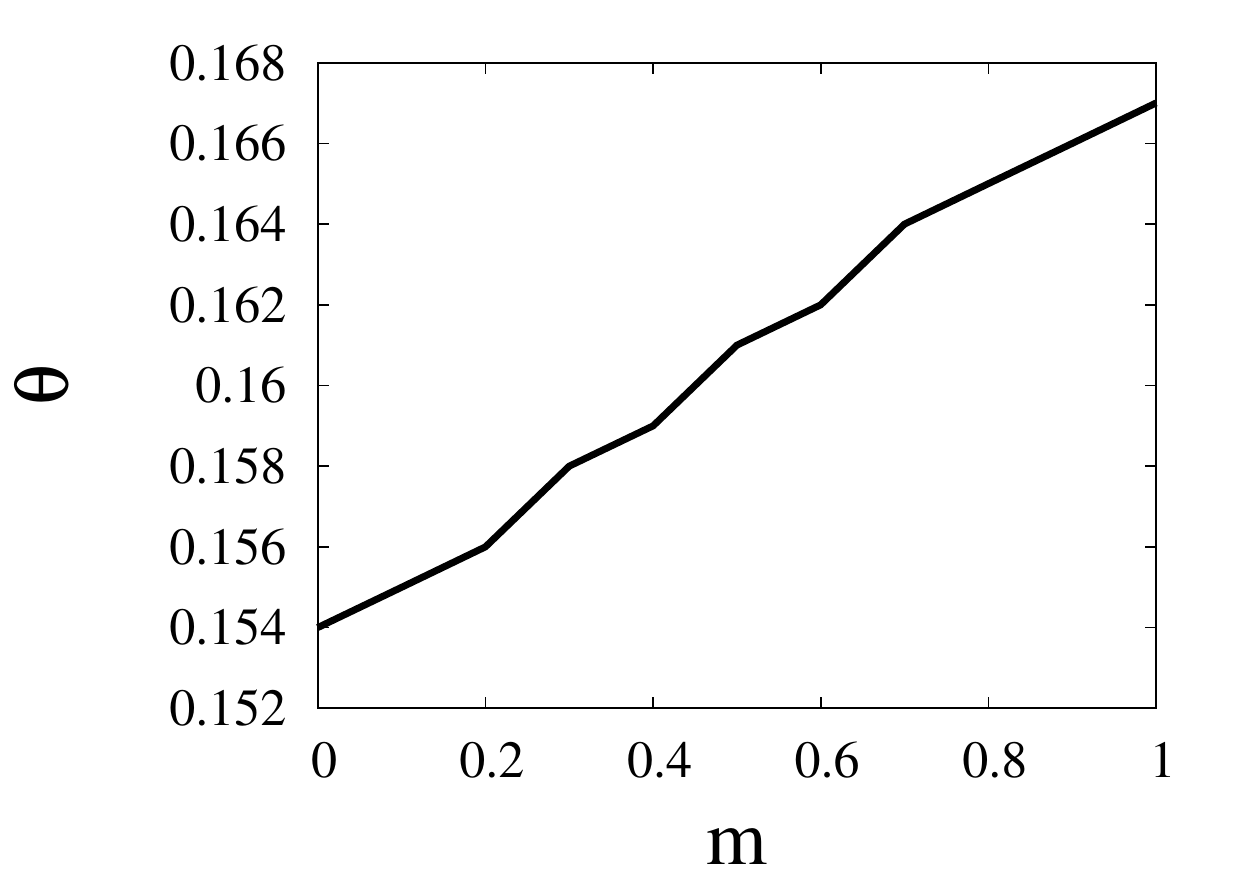}
               \caption{The $\theta$ which yields the \me rational map ansatz  \eqref{B=3-nonholo}.}
               \label{B=3-theta-vs-m} 
       \end{subfigure}%
        ~ 
        \begin{subfigure}[b]{0.5\textwidth}
                \centering \includegraphics[width=\textwidth]{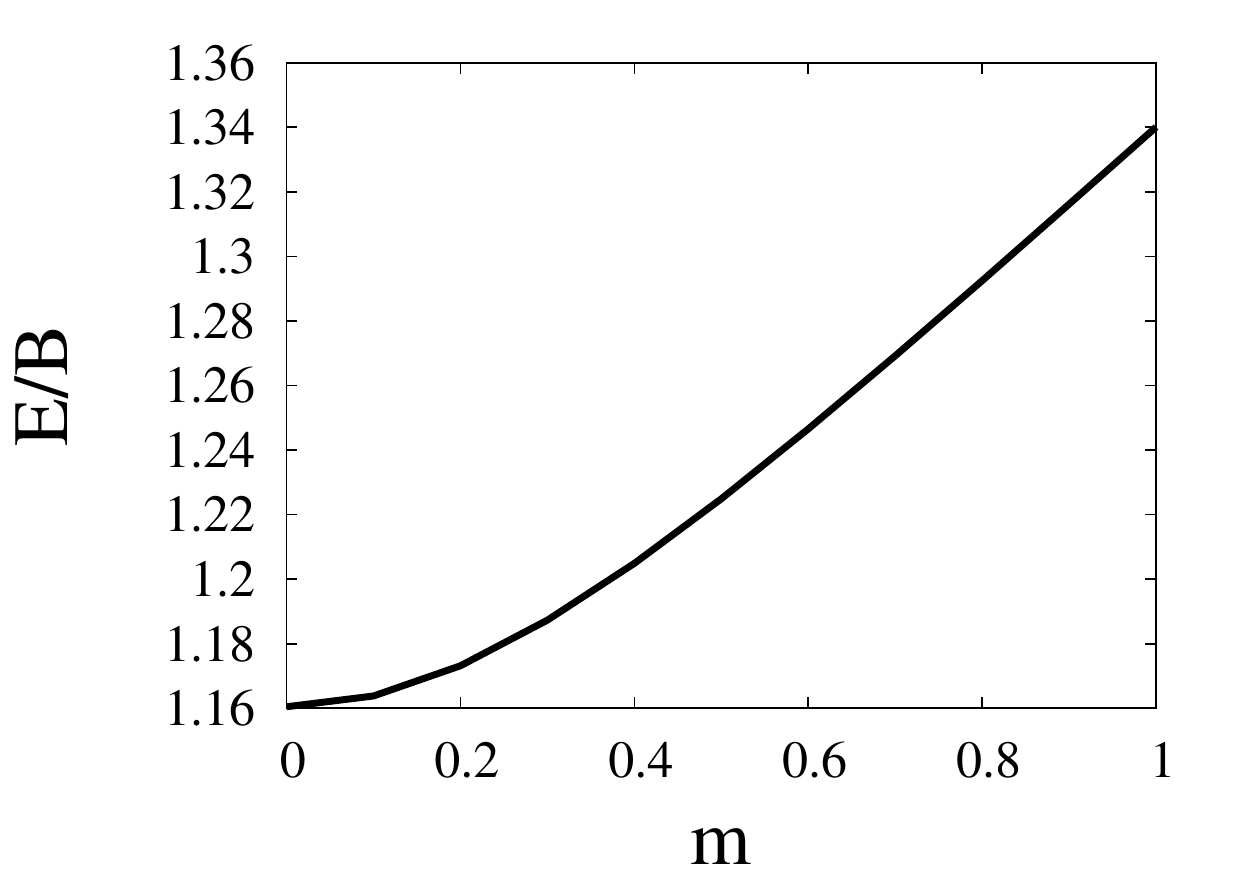}
                \caption{$E/B$ in \eqref{nonholo-E-rat-map} for the optimal value of $\theta$ as a function of $m$.}
                \label{B=3-EdivB-vs-m}
        \end{subfigure}
      \centering
       \begin{subfigure}[b]{0.5\textwidth}
               \centering \includegraphics[width=\textwidth]{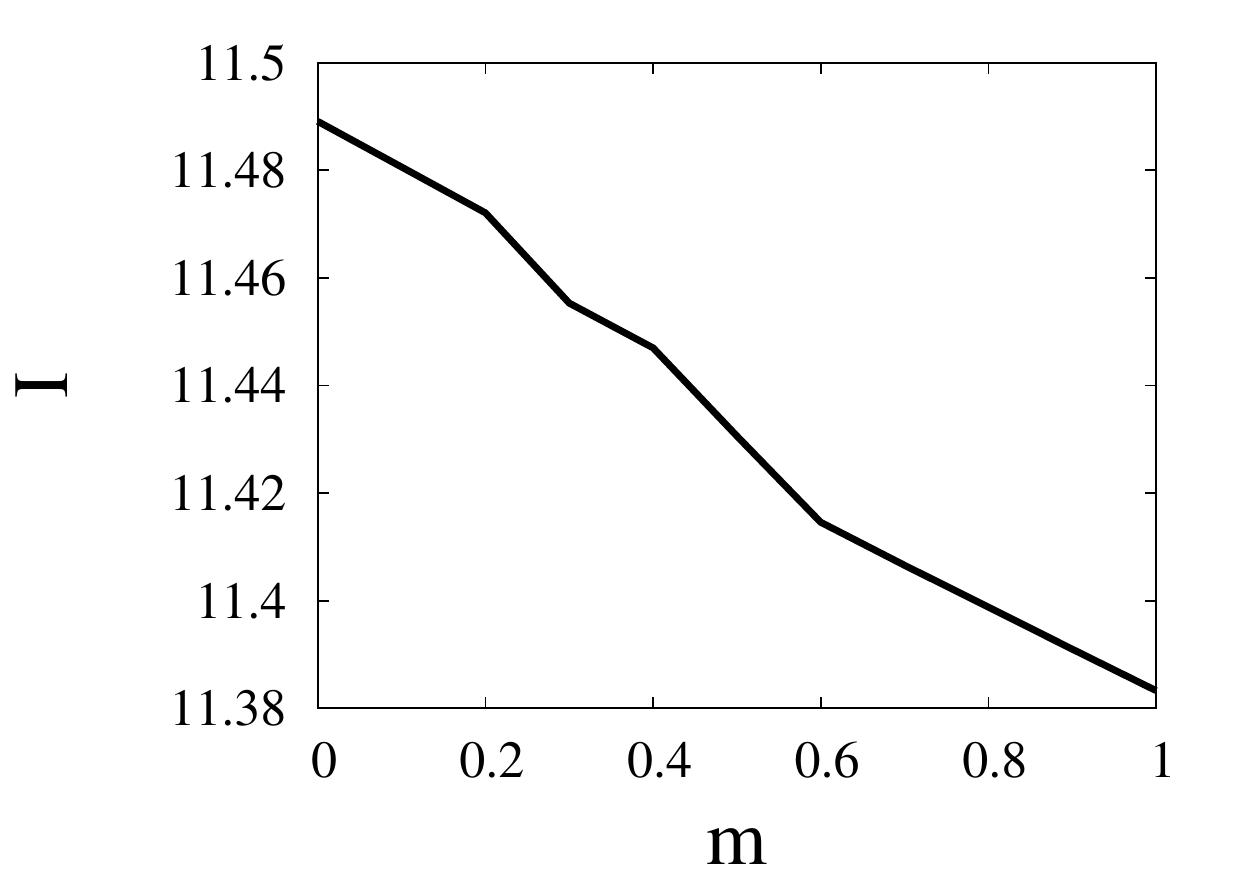}
               \caption{${\tilde{\cal I}}$ in \eqref{nonholo-I} as a function of $m$ for the optimal $\theta$.}
               \label{B=3-I-vs-m} 
       \end{subfigure}%
        ~ 
        \begin{subfigure}[b]{0.5\textwidth}
                \centering \includegraphics[width=\textwidth]{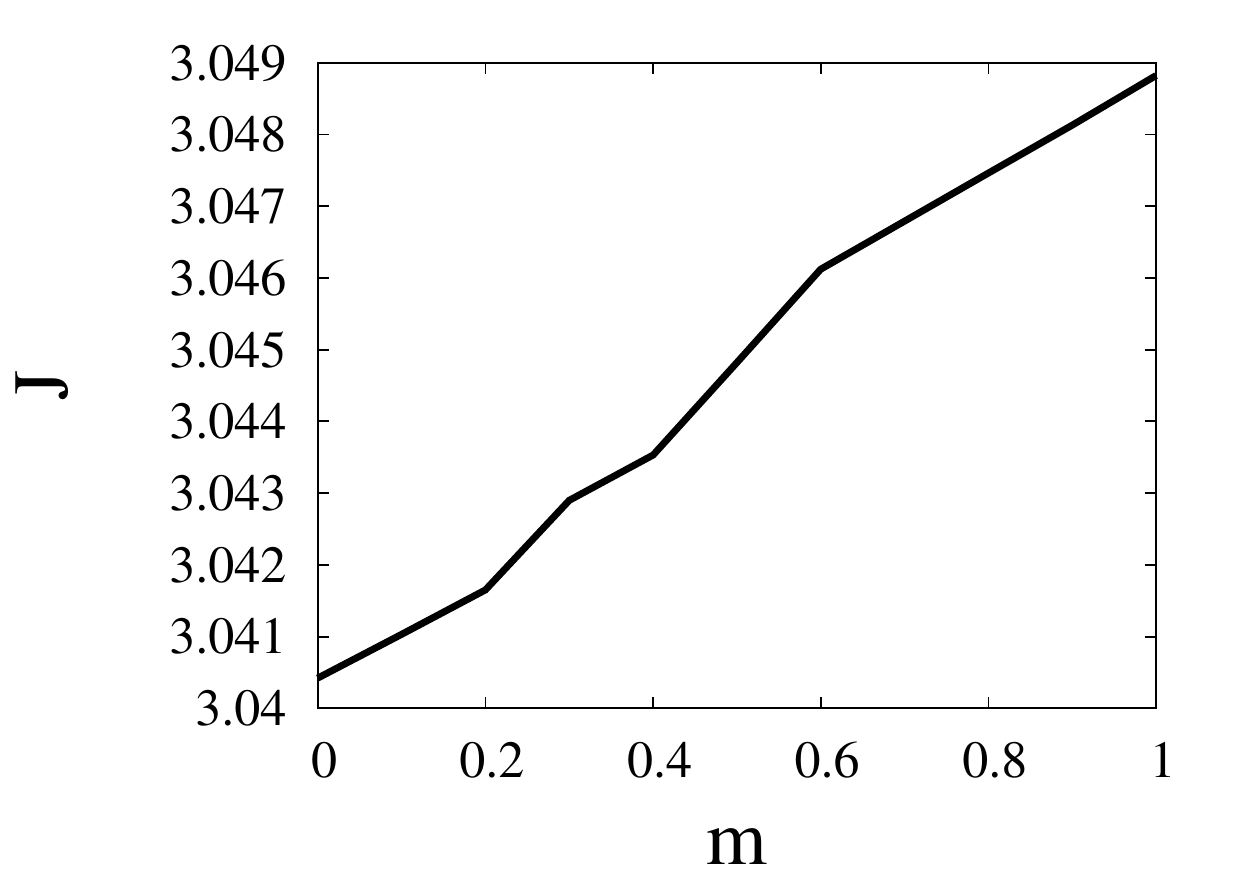}
                \caption{${\cal J}$ in \eqref{nonholo-I} as a function of $m$ for the optimal $\theta$.}
                \label{B=3-J-vs-m}
        \end{subfigure}
\caption{The \me $E/B$ and $\theta$ as a function of $m$ of \eqref{nonholo-E-rat-map}.}
\end{figure}

\subsection{The effects of the pion mass on the folding structure}
\label{pionmass}

It is well known that the Skyrme theory describes atomic nuclei better when a pion mass term is included, see \cite{Adkins:1983hy,Battye:2004rw, Battye:2009ad} for a discussion of $m \neq0$ and \cite{Battye:2005nx} for the implications of spinning Skyrmions. This inspired us to investigate how the surfaces of $\det(J(\bx))=0$ vary as a function of $m$. We first examine this question using the \nh rational map ansatz and then check our results using full field minimisation. As a side effect, this analysis provides a good test of the effectiveness of the rational map ansatz. 

In order to find the \me rational map in \eqref{nhratmap} for given $m$ we minimised the numerical integral of \eqref{nonholo-E-rat-map} with respect to $\theta$ using a standard search algorithm. The resulting function $\theta(m)$ is displayed in figure \ref{B=3-theta-vs-m} for values of $m$ between $0$ and $1$. The energy per baryon number is shown in figure \ref{B=3-EdivB-vs-m}. Note that $E/B$ increases monotonically as $m$ increases which agrees with the findings in \cite{Battye:2004rw}. A point worth noting is that it is energetically favourable for ${\tilde{\cal I}}$ to decrease slightly, see figure {\ref{B=3-I-vs-m}, and for ${\cal J}$ to increase as $m$ increases, see figure \ref{B=3-J-vs-m}. This has an effect on the behaviour of the shape function around the origin as can be seen from equation (\ref{f(r)=0}).

\begin{figure}[!htb]
\centering
       \begin{subfigure}[b]{0.5\textwidth}
               \centering \includegraphics[width=\textwidth]{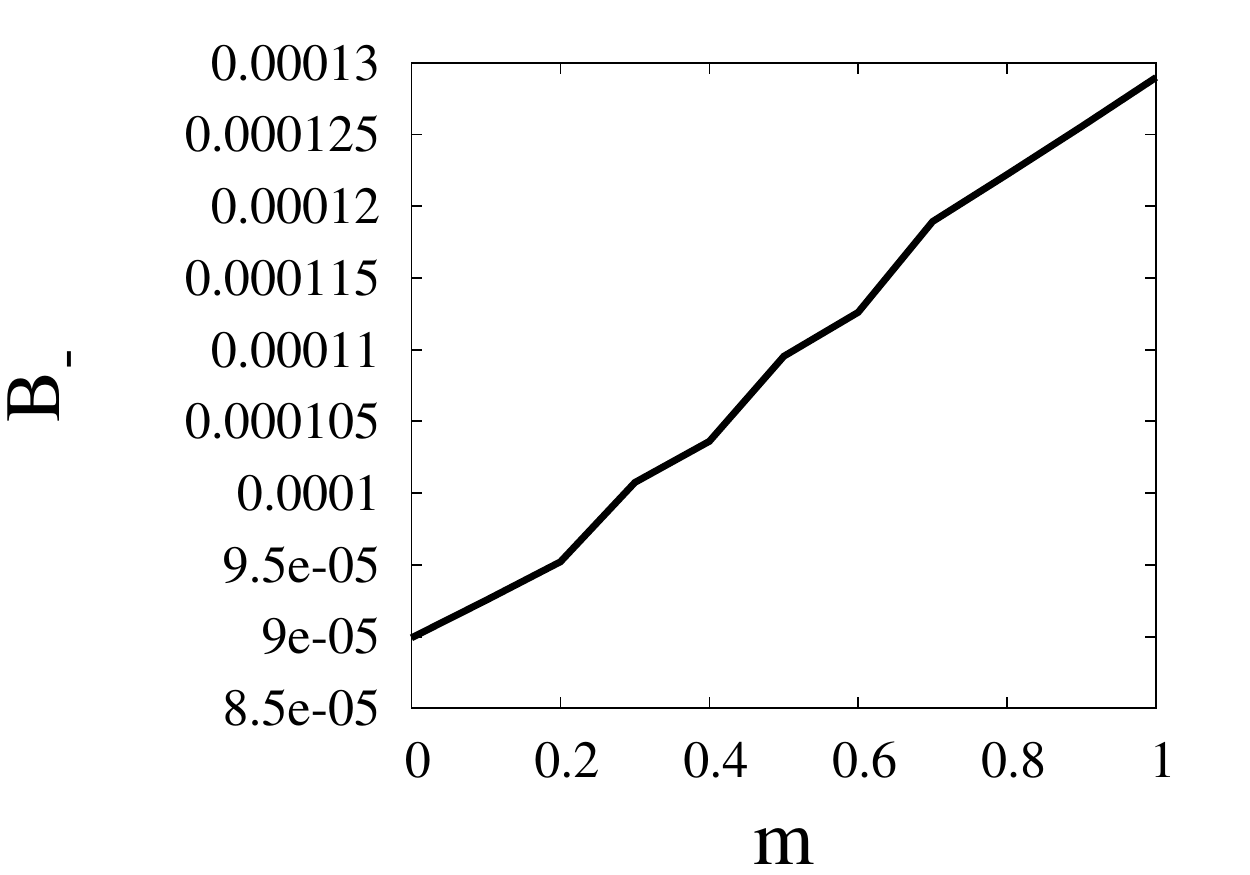}
               \caption{The total negative baryon density as a function of $m$.}
               \label{B=3-Neg-baryon-vs-m} 
       \end{subfigure}%
        ~ 
        \begin{subfigure}[b]{0.5\textwidth}
                \centering \includegraphics[width=\textwidth]{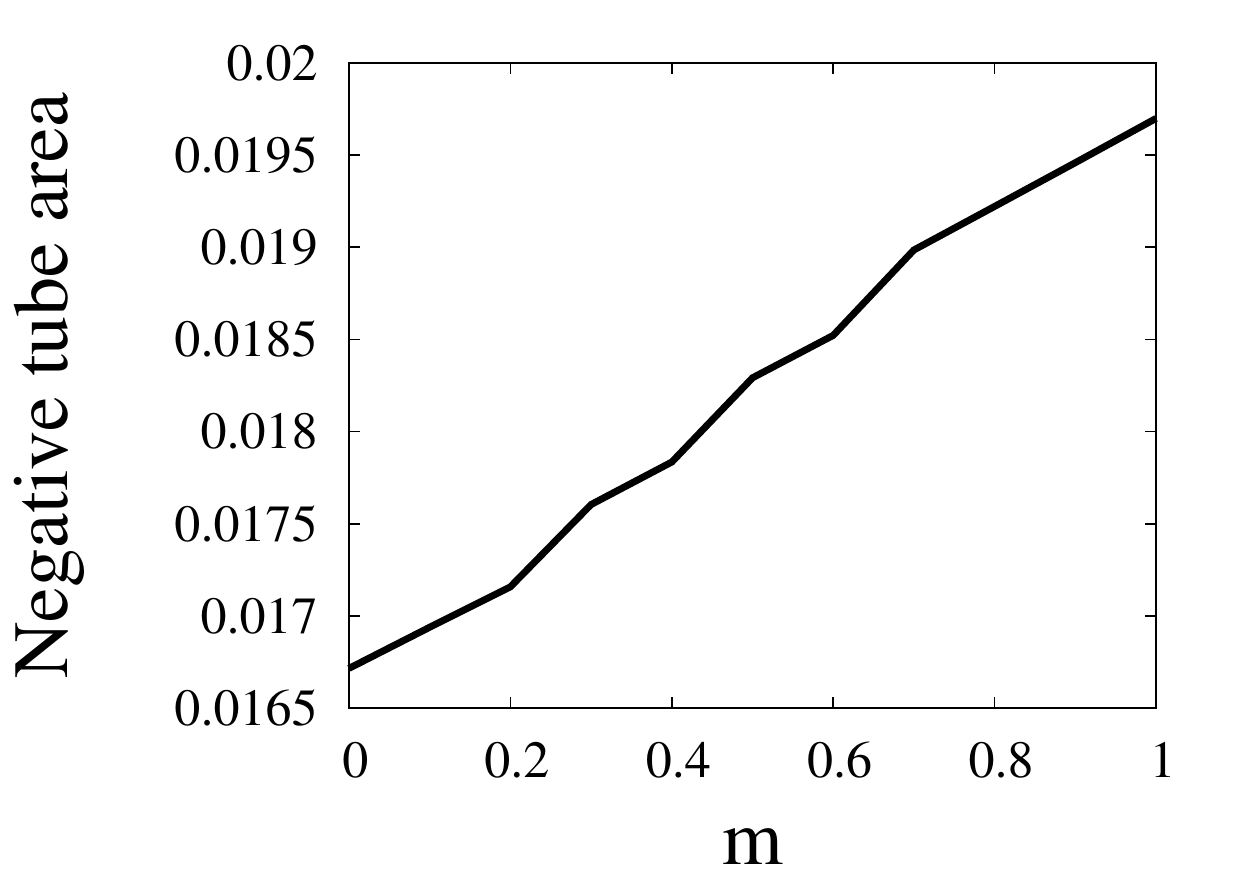}
                \caption{Area of the negative baryon density tubes as a function of $m$.}
                \label{B=3-Neg-tube-area-vs-m}
        \end{subfigure}
        \caption{Data arising from the \me $B=3$ rational map \eqref{B=3-nonholo}.}
\end{figure}

The radial integral of the baryon density in \eqref{nonholo-B} can be evaluated exactly using the boundary conditions of $f(r).$
Using the \me $\theta(m)$ we then integrated the negative baryon density ${\cal B}_-$ over $S^2$ to give the total negative baryon density $B_-$ as a function of $m$ as displayed in figure \ref{B=3-Neg-baryon-vs-m}. It is apparent that as the mass $m$ increases it is more energetically efficient for the $B=3$ Skyrmion to have more negative baryon density, $B_-$. 

In the  rational map ansatz, the negative baryon density arise as tubes emanating from the faces of the tetrahedral polyhedron, very similar to the yellow tubes in figure \ref{B3 negative baryon density}. To understand how these tubes change as a function of mass, we numerically integrated the area of a negative baryon density tube over $S^2$. This is shown in figure \ref{B=3-Neg-tube-area-vs-m}. Hence, the rational map ansatz predicts that the tubes of negative baryon density increase in size roughly linearly with $m.$

\begin{figure}[!htb]
       \centering
       \begin{subfigure}[b]{0.5\textwidth}
               \centering
\includegraphics[width=\textwidth]{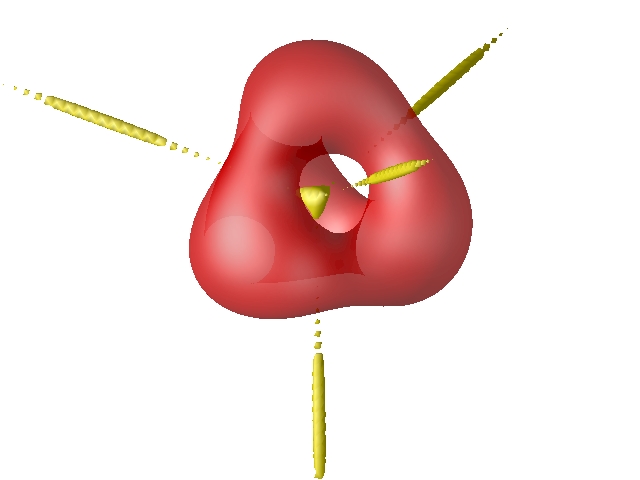}
               \caption{$m=0.1$}
               \label{B=3-m=0pt1} 
       \end{subfigure}%
        ~ 
        \begin{subfigure}[b]{0.5\textwidth}
                \centering
\includegraphics[width=\textwidth]{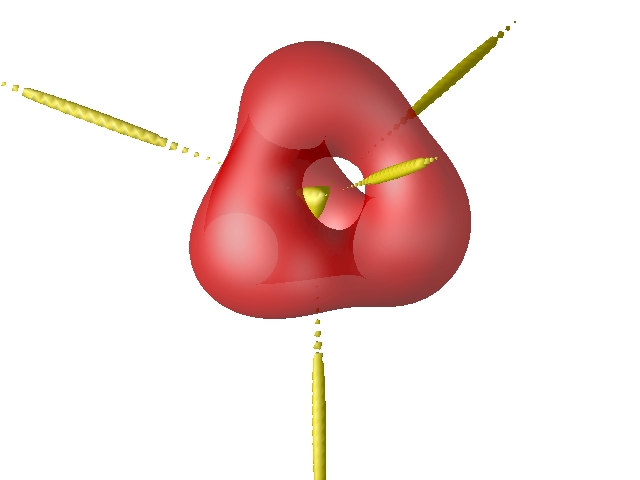}
                \caption{$m=0.2$}
                \label{B=3-m=0pt2}
        \end{subfigure}
\begin{subfigure}[b]{0.5\textwidth}
               \centering
\includegraphics[width=\textwidth]{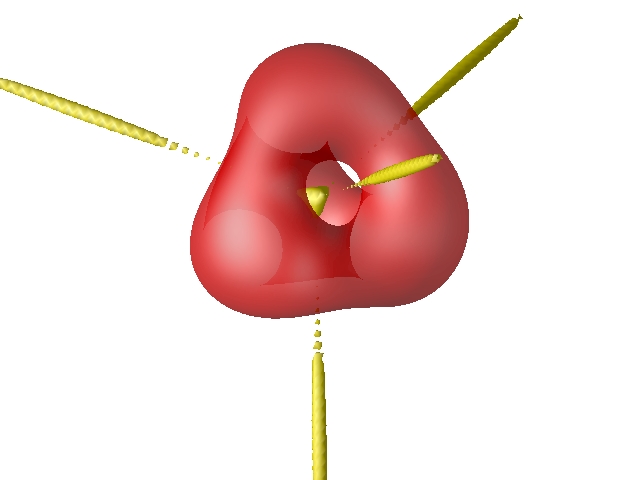}
               \caption{$m=0.3$}
               \label{B=3-m=0pt3} 
       \end{subfigure}%
        ~ 
        \begin{subfigure}[b]{0.5\textwidth}
                \centering
\includegraphics[width=\textwidth]{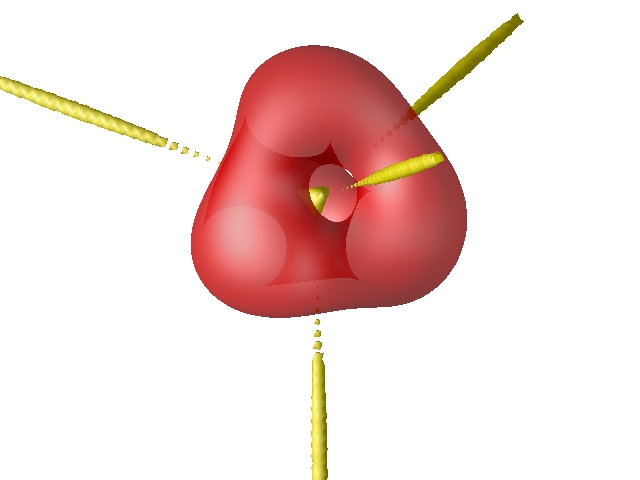}
                \caption{$m=0.4$}
                \label{B=3-m=0pt4}
        \end{subfigure}
     \begin{subfigure}[b]{0.5\textwidth}
               \centering 
\includegraphics[width=\textwidth]{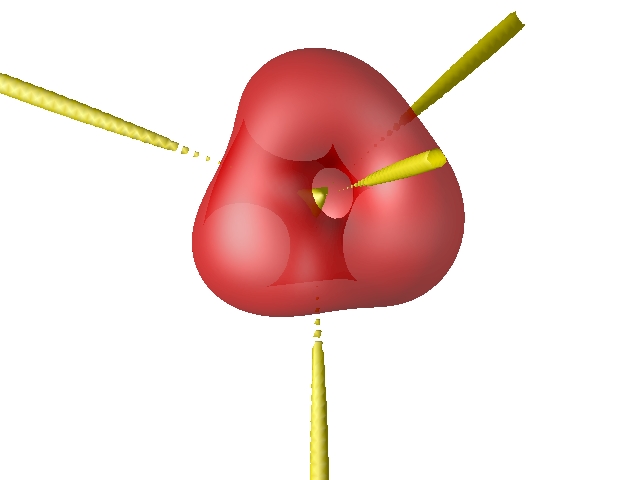}
               \caption{$m=0.5$}
               \label{B=3-m=0pt5} 
       \end{subfigure}%
        ~ 
        \begin{subfigure}[b]{0.5\textwidth}
                \centering
\includegraphics[width=\textwidth]{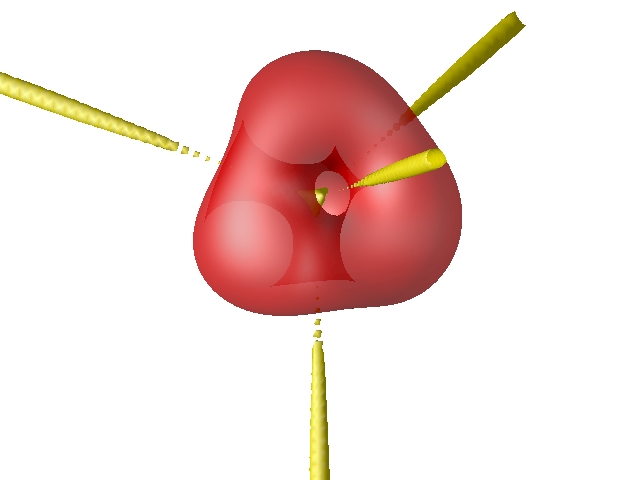}
                \caption{$m=0.6$}
                \label{B=3-m=0pt6}
        \end{subfigure}
\end{figure}

\begin{figure}[!htb]
\centering
\begin{subfigure}[b]{0.5\textwidth}
               \centering
\includegraphics[width=\textwidth]{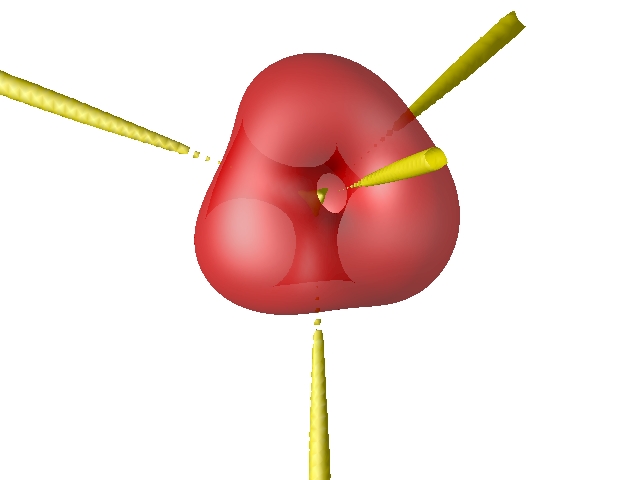}
               \caption{$m=0.7$}
               \label{B=3-m=0pt7} 
       \end{subfigure}%
        ~ 
        \begin{subfigure}[b]{0.5\textwidth}
                \centering
\includegraphics[width=\textwidth]{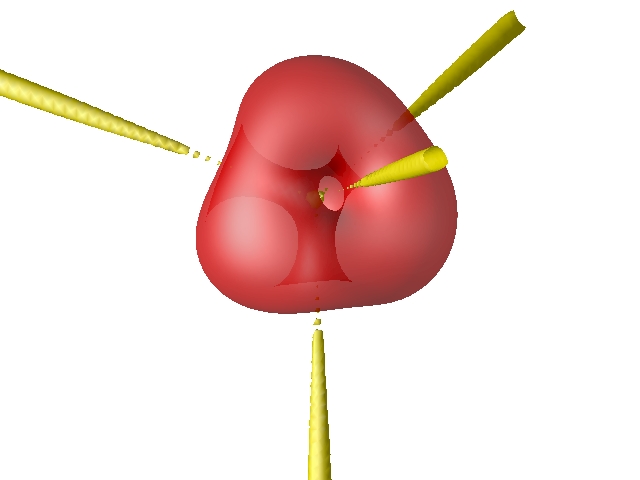}
                \caption{$m=0.8$}
                \label{B=3-m=0pt8}
        \end{subfigure}
\begin{subfigure}[b]{0.5\textwidth}
               \centering               \includegraphics[width=\textwidth]{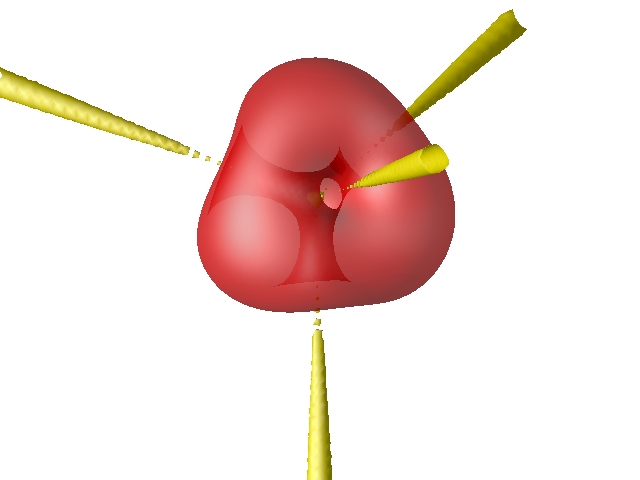}
               \caption{$m=0.9$}
               \label{B=3-m=0pt9} 
       \end{subfigure}%
        ~ 
        \begin{subfigure}[b]{0.5\textwidth}
                \centering
                \includegraphics[width=\textwidth]{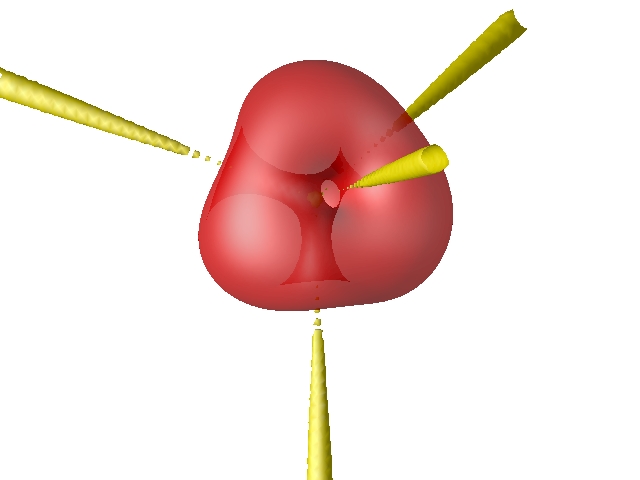}
                \caption{$m=1$}
                \label{B=3-m=1}
        \end{subfigure}
        \caption{$B=3$ Skyrmion of varying $m$. Red is a level set of $B=\mbox{const} >0$, yellow is $B=0$.}
\label{B=3 images}
\end{figure}

To proceed we numerically minimised the rational maps for the values of the pion mass given above using a full field minimisation. The results are shown in figures \ref{B=3 images}.
It is worth discussing the actual full field \me $B=3$ solutions. One should first note that, as expected and conventionally understood, the hole in the level-set of positive $\det\left(J(x)\right)$ becomes smaller for increasing $m$. This is well understood as the potential term forces the Skyrmion field to reach the vacuum value  exponentially. Also, as $m$ increases the singular tubes become more pronounced. This is clearly seen in figure \ref{B=3 images} where the singular tubes are much more defined for $m=1$ than for $m=0.1$.

\begin{figure}[!htb]
\centering
\begin{subfigure}[b]{0.5\textwidth}
               \centering
               \includegraphics[width=\textwidth]{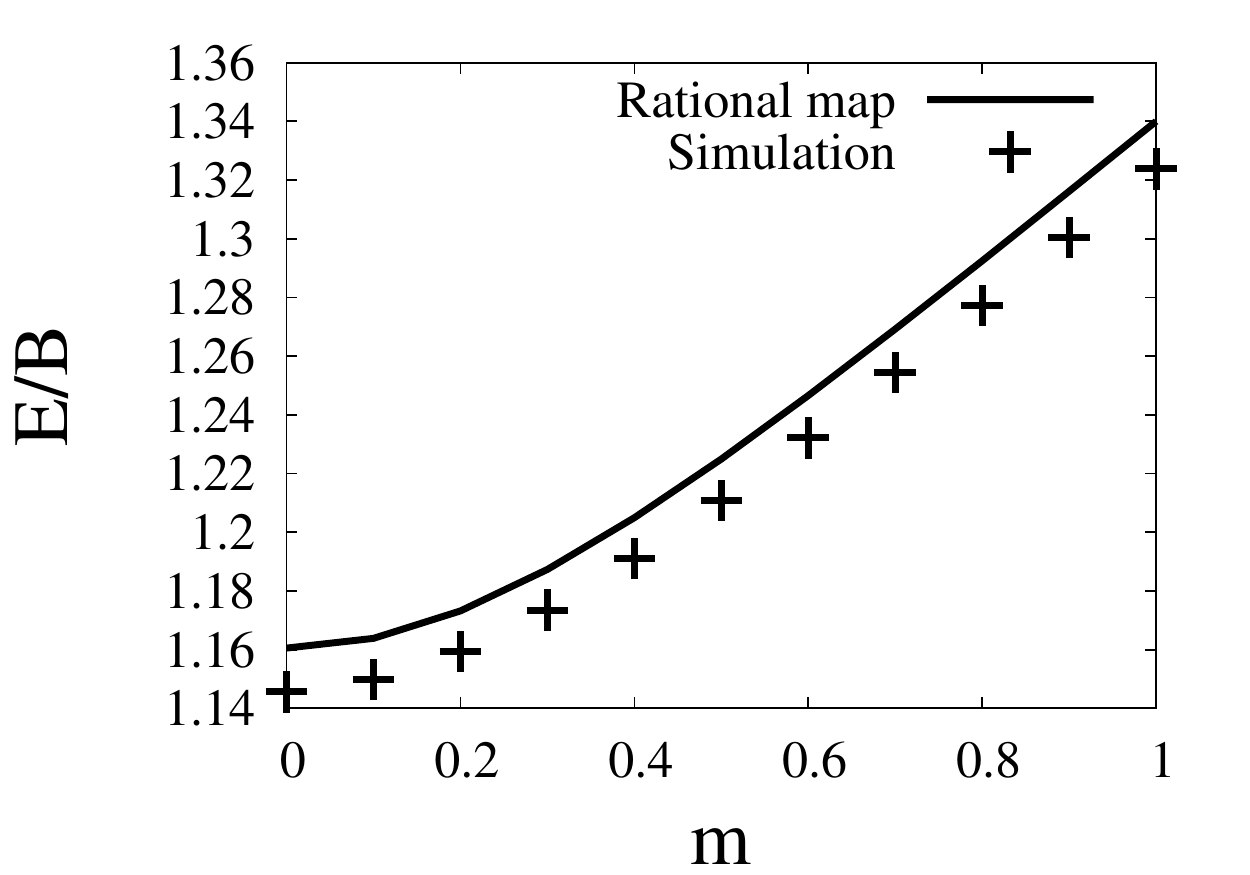}
               \caption{Comparison of $E/B$ for the \nh rational map ansatz and $E/B$ found from full field minimisation as a function of $m$.}
               \label{EdivB-Numerical-rational-map} 
       \end{subfigure}%
        ~ 
        \begin{subfigure}[b]{0.5\textwidth}
                \centering
\includegraphics[width=\textwidth]{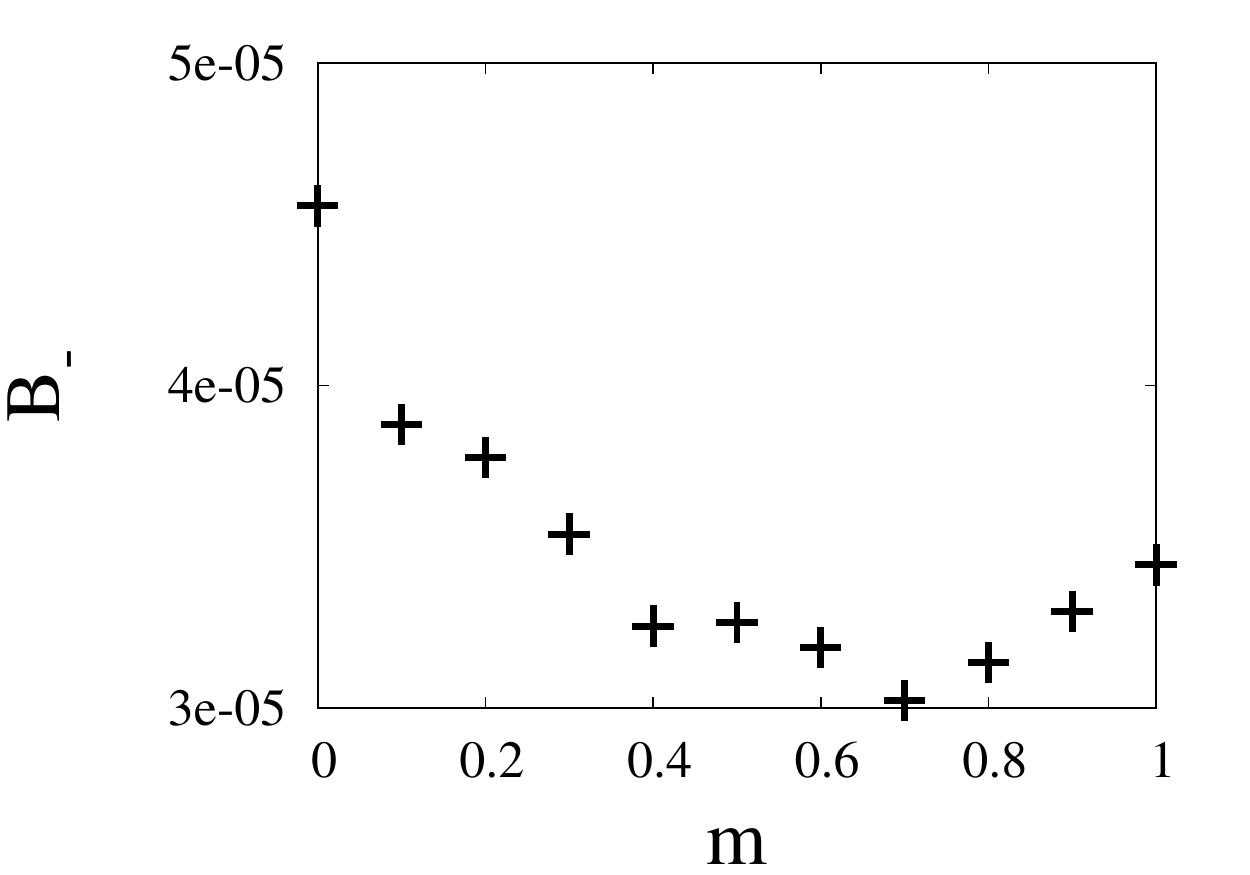}
                \caption{Negative baryon density in full field \me solutions}
                \label{Neg-B-in-B=3}
        \end{subfigure}
\begin{subfigure}[b]{0.5\textwidth}
               \centering \includegraphics[width=\textwidth]{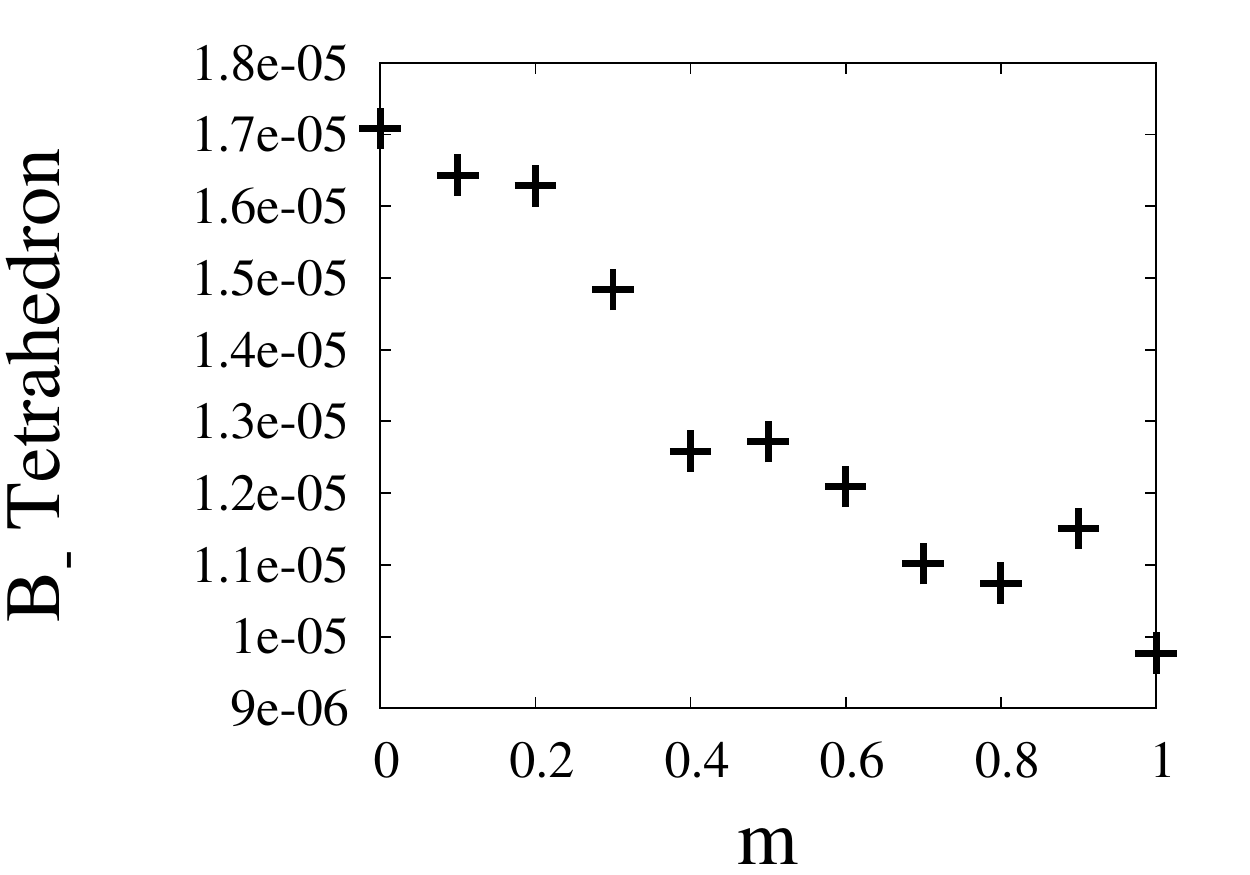}
               \caption{Negative baryon density in the central tetrahedron as a function of $m$.}
               \label{B=3-neg-tetra} 
       \end{subfigure}%
        ~ 
        \caption{$B=3$ full field \me solutions data.}
\end{figure}

 As shown in figure \ref{EdivB-Numerical-rational-map}, the energy per baryon for the full field minimisation is about $2\%$ lower than that for the rational map ansatz, for all $m$. This is expected, because as previously shown the rational map ansatz forces all of the pre-images of the anti-vacuum to be at the origin. 

As can be seen in figure \ref{B=3-neg-tetrahedron} the Skyrme field has a fifth pre-image at the origin with negative orientation showing that there is also negative baryon density at the origin. This region of negative baryon density is inside a tetrahedron, which is dual to a surface of positive baryon density. In this central region $\sigma \approx -1$, this is where the potential energy (the pion mass term) is maximal hence as $m$ increase the negative baryon density at the origin decreases. To verify this we numerically integrated over the central tetrahedron of negative baryon density, this is shown in figure \ref{B=3-neg-tetra}. This shows that indeed as $m$ increases the total negative baryon density in the central dual tetrahedron decreases. This gives a heuristic explanation why the total negative baryon density initially decreases, shown in figure \ref{Neg-B-in-B=3}, as $m$ starts to increase from $m=0$, then after $m \approx 0.7$ the negative baryon density starts to increase again. This is reflected in the \nh rational map ansatz. This trend is further verified because we also found the \me $B=3$, $m=2$ solution. This solution was found to have $E/B=1.551$,
$B_-=4.1 \times 10^{-5}$ and a $B_-$ dual tetrahedron total baryon density of $1.2 \times 10^{-5}$. Figure \ref{B=3-neg-tetra} is not very smooth. This is most likely due to the numerical grid and the algorithm which identified the edge of the dual tetrahedron.

\subsection{Expansion around the origin}
\label{expansion}

The tetrahedral symmetry of the $B=3$ Skyrme field poses stringent restrictions on the allowed terms in a Taylor expansion around the origin. Here we calculate the allowed polynomials and estimate the relevant coefficients from the numerical solution. 

The rational map of the $B=3$ Skyrmion is given by 
\begin{equation}
R(z) = \frac{\sqrt{3} i z^2 - 1}{z^3 -\sqrt{3} i z}
\end{equation}
which is $T_d$ tetrahedrally symmetric using the same orientation as in \cite{Houghton:1997kg}. This symmetry is generated by a $C_2$
symmetry
\begin{equation}
z \mapsto -z, \quad R \mapsto -R,
\end{equation}
and a $C_3$ symmetry
\begin{equation}
z \mapsto \frac{iz+1}{-iz+1},
\quad R \mapsto 
\frac{-iR+i}{R+1},
\end{equation}
for the tetrahedral symmetry $T$, together with an additional reflection symmetry
\begin{equation}
z \mapsto \frac{{\bar z}-i}{-i{\bar z}+1},
\quad R \mapsto 
\frac{{\bar R}-i}{-i{\bar R}+1}.
\end{equation}

The corresponding symmetries in component notation for the $C_2$
generator is
\begin{equation}
\left(x_1,x_2,x_3\right) \mapsto \left(-x_1,-x_2,x_3\right), 
\quad \left(\pi_1,\pi_2,\pi_3\right)
\mapsto\left(-\pi_1,-\pi_2, \pi_3\right).
\end{equation}
and
\begin{equation}
\left(x_1,x_2,x_3\right) \mapsto \left(x_2,x_3,x_1\right),\quad
\left(\pi_1,\pi_2,\pi_3\right) 
\mapsto \left(\pi_3, \pi_1, \pi_2\right)
\end{equation}
for the $C_3$ generator. The reflection symmetry is given by
\begin{equation}
\left(x_1,x_2,x_3\right) \mapsto \left(x_1,-x_3,-x_2\right),\quad
\left(\pi_1,\pi_2,\pi_3\right) 
\mapsto \left(\pi_1, -\pi_3, -\pi_2\right)
\end{equation}

Hence the field ${\pmb \pi}$ can be expanded
around the origin as
\begin{eqnarray}
\nonumber
{\pmb \pi} &=& 
a_1 \left(
\begin{array}{c}
x \\ y \\ z
\end{array}
\right)
+
b_1 \left(
\begin{array}{c}
yz \\ xz \\ xy
\end{array}
\right)
+ c_1 \left(
\begin{array}{c}
x^3 \\y^3 \\ z^3
\end{array}
\right)
+ c_2 r^2\left(
\begin{array}{c}
x \\y \\ z
\end{array}
\right)
\\
\label{expansion}
&&{}+
d_1 \left(
\begin{array}{c}
yzx^2 \\  xzy^2\\xyz^2
\end{array}
\right)
+ d_2 r^2 \left(
\begin{array}{c}
y z \\  x z \\ x y 
\end{array}
\right)
+\dots,
\end{eqnarray}
where the Cartesian coordinates are now denoted as $(x,y,z)$ and $r^2=x^2+y^2+z^2.$ The tetrahedral symmetry can be augmented to spherical symmetry by setting $b_1=0,$ $c_1=0$ and $d_1=d_2=0,$ which corresponds
to the Taylor expansion of the hedgehog ansatz around the origin. This provides a useful check that we have implemented the tetrahedral symmetry correctly.

\begin{figure}[!htb]
\centering
\includegraphics[height=3.8in]{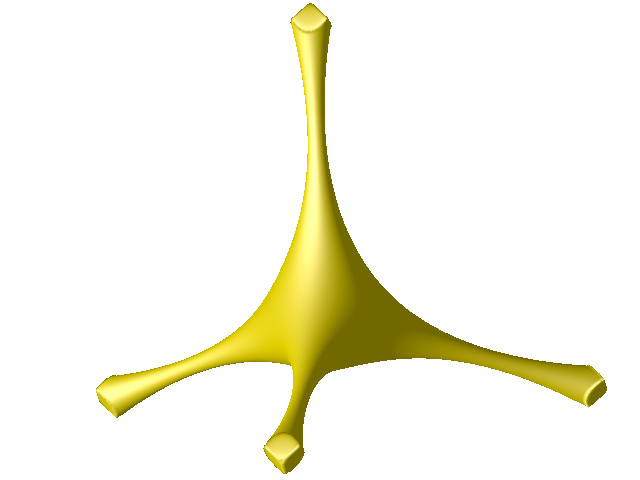} 
\caption{Singular surface $\det J = 0$ calculated using the polynomial expansion around the origin in (\ref{expansion}).}
\label{J0expansion}
\end{figure}

In order to compare the Taylor expansion of ${\pmb \pi}$ to the numerical solution, we have to calculate the coefficients in \eqref{expansion}. We used the following approach. Setting $y=0$ and $z=0$ in \eqref{expansion} gives a polynomial in $x.$ We fitted this polynomial to our numerical data using a least square fit, and this allows us to calculate $a_1$ and $c_1+c_2.$ Similarly, setting $y=x,$ and $z=0$ gives $a_1,$ $b_1,$ $c_1+2c_2$ and $2d_2.$ Finally, setting $y=x$ and $z=x$ also gives an equation in $d_1.$ In order to improve the approximation close to the origin, we fitted to a polynomial of higher degree. By plotting numerical data and approximation we found that the values $a_1=-0.41,$ $b_1 = -2.1,$ $c_1 = -2.2,$ $c_2=1.2,$ $d_1=0.76$ and $d_2 = 0.64$ are a reasonable approximation for $-0.5 < x,y,z < 0.5.$ Note that the errors in particular for the coefficients $d_1$ and $d_2$ are rather large. 
We can now evaluate the Jacobian of the map. At the origin ${\bf
  x}=0,$ the Jacobian is non-zero, namely, $\det(J_0) = a_1^3 < 0.$ The value of $\det J_0(0)= -0.055$ calculated numerically from the exact solution matches the value for the expansion \eqref{expansion}.

Figure \ref{J0expansion} shows a plot of the surface $\det J=0$ inside a cube of length $0.5$. This clearly looks very similar to the surface $\det J=0$ arising from the numerical solution displayed in figure \ref{B=3 sigma =1}. 
Using our expansion, we can check whether the singular surface pinches off at a point. By setting $y=x,$ and $z=x$ in the Jacobian $J_0$ we can show that $\det J_0 = 0$ for $x = -0.09$ but there is no positive solution within the box. We were careful to include terms up to fourth order because the normal form of the swallowtail includes a fourth order term. In summary, we have deduced the following folding structure for the $B=3$ Skyrmion. There are four folding tubes through the faces of a tetrahedron. These tubes smoothly connect to the corners of a dual tetrahedron at the origin. The folding surface are not intersecting each other, so there are no swallowtail singuarities in the $B=3$ Skyrmion, contrary to the conjecture in \cite{Houghton:2001fe}. By symmetry, the cusp lines are expected to lie on the edges of the tetrahedron, and there is some numerical evidence.

\section{Conclusion}
This paper was motivated by the results of \cite{Houghton:2001fe}, where the authors found regions of negative baryon density in the rational maps for the $B=3$ \me Skyrmion. For small $m$ these regions of negative baryon density are very small, but we have been able to numerically verify their existence. Also we have discovered a tetrahedron of negative baryon density at the origin, which is dual to a tetrahedron produced as a level-surface of positive constant baryon density. The singularities corresponding to surfaces of zero Baryon density form four tubes which smoothly join up at the dual tetrahedron at the origin. Contrary to the conjecture in \cite{Houghton:2001fe}, there are no swallowtail singularities in the $B=3$ Skyrmion configuration. We have also found that for the $B=3$ Skyrmion there are five pre-images of the anti-vacuum. Four with positive orientation, on the vertices of a tetrahedron, and one with negative orientation at the origin. This behaviour is also seen in monopoles \cite{Sutcliffe:1996he}. 

The authors of \cite{Houghton:2001fe} did not find any regions of negative baryon density in the rational map ansatz for the $B=4$ \me solution. This has been verified here numerically. Furthermore, assuming octahedral symmetry we have shown in appendix \label{B=4} that there are no regions of negative baryon density around the origin. These results are also consistent with the instanton ansatz \cite{Leese:1993mc}.

It has already been discussed that regions of negative Jacobian-determinant occur for the charge three instanton \cite{Leese:1993mc}. This is signifiant because there is a BPS extended Skyrme model which can be derived from Yang-Mills instantons \cite{Sutcliffe:2010et}, which must contain regions of negative baryon density. This extended model has an infinite number of vector mesons and rho mesons. There has also been research into a truncated version of this model \cite{Sutcliffe:2011ig}, where only a few extra terms are included. Understanding the form and distribution of the negative baryon density in these models would be very interesting. It should be noted that there is another BPS Skyrme model \cite{Adam:2012sa}, where the Bogomolny equation shows that the baryon density is proportional to the square root of the potential. Hence, if the potential is positive definite through space, so is the baryon density \footnote{This was pointed out by M. Speight.}.

\section*{Acknowledgements}

SK would like to thank N Manton for useful discussions, and in particular for discussing the singularities of $B=4.$ The authors acknowledge the EPSRC for the grant EP/I034491/1.

\appendix
\section{$B=4$}
\label{B=4}
It has been shown \cite{Houghton:2001fe} that even when the $B=4$ rational map ansatz is extended to be \nh no negative baryon density is found. This is also seen in our full field minimisation of the $B=4$ Skyrmion for $m=0$ and $m=1,$ see figure \ref{B=4m=0} and \ref{B=4m=1}, respectively. For $m=0$ we found $E/B=1.12$, and for $m=1$ we found $E/B=1.30$. In both
cases, up to numerical accuracy, we did not find any negative baryon density.

\begin{figure}[H]
       \centering
       \begin{subfigure}[b]{0.5\textwidth}
               \centering
               \includegraphics[width=\textwidth]{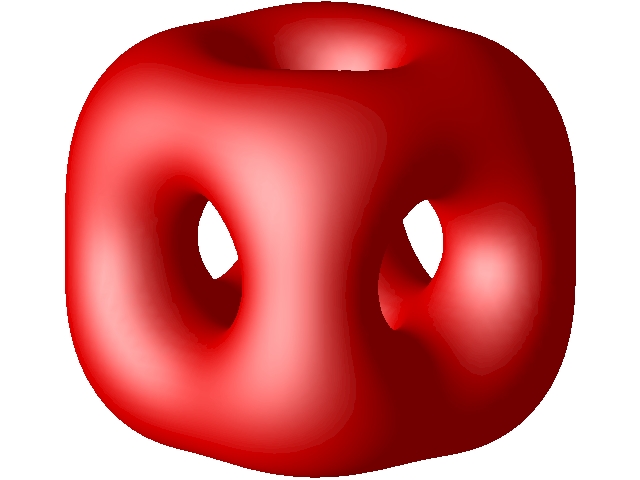}
               \caption{$m=0$}
               \label{B=4m=0} 
       \end{subfigure}%
        ~ 
        \begin{subfigure}[b]{0.5\textwidth}
                \centering
                \includegraphics[width=\textwidth]{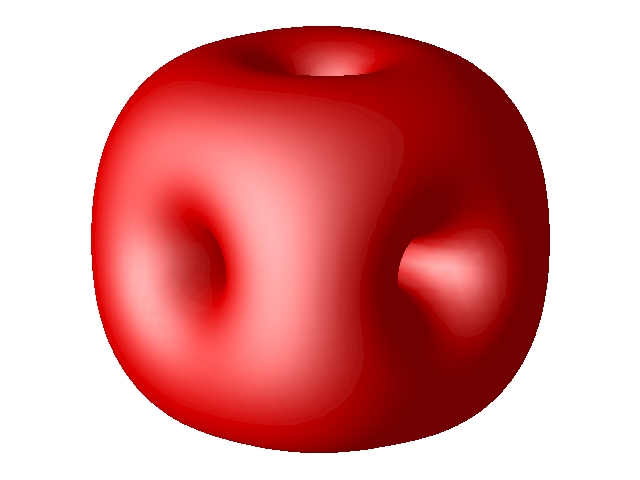}
                \caption{$m=1$}
                \label{B=4m=1}
        \end{subfigure}
        ~ 

        \caption{\me full field solutions of the $B=4$. Red is a surface of constant positive $\det(J)$.}
\end{figure}

In the following, we show that the $B=4$ \me octahedrally symmetric Skyrmion
does not have regions of negative baryon density near the origin. This is where the \nh rational map ansatz is not a good approximation of the exact solution. As a starting point, we consider the rational map of the $B=4$ Skyrmion is given by
\begin{equation}
R(z) = \frac{z^4+2\sqrt{3} i z^2 + 1}{z^4 - 2 \sqrt{3} i z^2 + 1}
\end{equation}
which is octahedrally symmetric. This symmetry is generated by a $C_4$
symmetry:
\begin{equation}
z \mapsto i z, \quad R \mapsto \frac{1}{R},
\end{equation}
and a $C_3$ symmetry
\begin{equation}
z \mapsto \frac{iz+1}{-iz+1},\quad R \mapsto {\rm e}^{-\frac{2\pi
    i}{3}} R.
\end{equation}
The corresponding symmetries in component notation for the $C_4$
generator is
\begin{equation}
\left(x_1,x_2,x_3\right) \mapsto \left(x_2,-x_1,x_3\right), 
\quad \left(\pi_1,\pi_2,\pi_3\right)
\mapsto\left(\pi_1,-\pi_2, -\pi_3\right).
\end{equation}
and
\begin{equation}
\left(x_1,x_2,x_3\right) \mapsto \left(x_2,x_3,x_1\right),\quad
\left(\pi_1,\pi_2,\pi_3\right) 
\mapsto \left(-\frac{1}{2} \pi_1 - \frac{1}{2} \sqrt{3} \pi_2, 
\frac{1}{2}\sqrt{3} \pi_1 - \frac{1}{2} \pi_2,
  \pi_3\right)
\end{equation}
for the $C_3$ generator. Hence the field ${\pmb \pi}$ can be expanded
around the origin as
\begin{eqnarray*}
{\pmb \pi} &=& 
b_1 \left(
\begin{array}{c}
x^2+y^2-2z^2 \\ -\sqrt{3}x^2 + \sqrt{3}y^2 \\0
\end{array}
\right)
+ c_1 \left(
\begin{array}{c}
0 \\0 \\ xyz
\end{array}
\right)\\
&&{}+d_1 \left(
\begin{array}{c}
x^4+y^4-2z^4 \\ -\sqrt{3}x^4 + \sqrt{3}y^4 \\0
\end{array}
\right)
+ d_2 \left(
\begin{array}{c}
2x^2y^2-x^2z^2-y^2z^2 \\ -\sqrt{3}x^2z^2 + \sqrt{3}y^2z^2 \\0
\end{array}
\right)
+\dots
\end{eqnarray*} 
A similar expansion as $r$ tends to infinity has been performed in \cite{Feist:2011aa}. We can now evaluate the Jacobian of the map. At the origin ${\bf  x}=0,$ the Jacobian is identically zero. Hence the singularity is
  clearly non-generic since for a generic singularity the Jacobian has
  rank $2.$ The determinant of the Jacobian can be expanded into
  $O-$symmetric polynomials. The lowest order term is
$$
\det(J) = \sqrt{3} b_1^2c_1 \left(x^2y^2+y^2z^2+z^2x^2\right) + \dots
$$
Hence the singularities are on the three coordinate axis which meet at
the origin. Furthermore, sufficiently close to the origin $\det(J)$ is positive, and hence there is no negative baryon density.

\bibliographystyle{utphys}
\bibliography{hopf2.bib}

\end{document}